\documentstyle[11pt]{article} 
\begin{document} 
\newcommand{\UGR}{\em Dpto F\'{\i}sica Te\'orica y del Cosmos\\ 
		       Universidad de Granada, 18071 Granada, Spain} 
\newcommand{\UAB}{\em Grup de F\'{\i}sica Te\`orica,\\ 
	      Universitat Aut\`onoma de Barcelona, 08193 Bellaterra, Spain}  
					      
\newcommand{\NPB}[3]{{ Nucl.\ Phys.\/} {\bf B#1} (19{#3}) {#2}}  
\newcommand{\PLB}[3]{{ Phys.\ Lett.\/} {\bf #1B} (19{#3}) {#2} } 
\newcommand{\PRD}[3]{{ Phys.\ Rev.\/}  {\bf D#1} (19{#3}) {#2} } 
\newcommand{\PRA}[3]{{ Phys.\ Rev.\/}  {\bf A#1} (19{#3}) {#2} } 
\newcommand{\PRL}[3]{{ Phys.\ Rev.\ Lett.\/} {\bf #1} (19{#3}) {#2} } 
\newcommand{\ZFP}[3]{{ Zeit.\ f.\ Phys.\/}, {\bf C #1} (19{#3}) {#2} } 
\newcommand{\IJA}[3]{{ Int.\ J.\ Mod.\ Phys.\/} {\bf A#1} (19{#3}) {#2} }

\def\lhs{{\it l.h.s.\/ }} 
\def\rhs{{\it r.h.s.\/ }} 
\def\bqn{\begin{equation}} 
\def\eqn{\end{equation}} 
\def\bqna{\begin{eqnarray}} 
\def\eqna{\end{eqnarray}} 
\def\nn{\nonumber} 
\def\bit{\bibitem} 
\def\vs{\vspace{.25in}} 
\def\thefootnote{\fnsymbol{footnote}} 
 
\def\QQa{\renewcommand{\baselinestretch}{1.3}\Huge\large\normalsize} 
 
\makeatletter 
\def\secteqno{\@addtoreset{equation}{section}%
\def\theequation{\thesection.\arabic{equation}}} 
 
\def\endsecteqno{\def{theequation\{\@ifundefined{chapter}%
{\arabic{equation}}{\thechapter.\arabic{equation}}}} 
\makeatother 
 
\def\ks{K_S} 
\def\kl{K_L} 
\def\k0{K^0} 
\def\kb0{\bar{K^0}} 
\def\reaction{\pi^- p \to \Lambda K^0} 
\def\jxt{\vec{j}(\vec{x},t)} 
\def\GeV{{\rm GeV}} 
\pagestyle{empty} 
{\hfill \parbox{6cm}{\begin{center} UAB-FT-393 \\ 
				    UGR-FT-64 \\ 
				    May 1996 
		      \end{center}}} 
	      
\vspace*{2cm}                               
\begin{center} 
\large{\bf SPACE-DEPENDENT PROBABILITIES FOR $K^0-\bar{K^0}$
OSCILLATIONS} 
\vskip .6truein 
\centerline {B. Ancochea$^a$, A. Bramon$^a$, R. Mu\~noz-Tapia$^b$  
	     and M. Nowakowski$^a$}  
\end{center} 
\vspace{.3cm} 
\begin{center} 
$^a${\UAB} \\ 
$^b${\UGR} 
\end{center} 
\vspace{1.5cm} 
  
\centerline{\bf Abstract} 
\medskip 
We analyze $K^0-\bar{K^0}$ oscillations in space in terms of propagating  
wave packets with coherent $K_S$ and $K_L$ components. 
The oscillation probabilities $P_{K^0 \to K^0} (x)$ 
and  $P_{K^0 \to \bar{K^0}} (x)$ depending only on the distance $x$, 
are defined through the time integration of a
current density $j(x,t)$ . The definition is such that it 
coincides with the experimental setting, thus avoiding 
some ambiguities and clarifying some controversies that have been 
discussed recently. 
 
\newpage

\pagestyle{plain} 
\QQa 
 
\section{Introduction} 
Due to ongoing and planned experiments in kaon physics there has 
been a renewed interest in space--time oscillations of neutral  
meson systems like $K^0$--$\bar{K^0}$. In particular, 
in \cite{1,2,3} an effort has been made to compute 
oscillation probabilities, like $P_{\k0 \to \k0} (r)$ and 
$P_{\k0 \to \kb0} (r)$,  which depend now solely on the  
spatial distance $r$ between the $\k0$ production point 
and the position where $\k0$ or $\kb0$ are detected. 
The expression $P(r)$ mimics then closely the usual experimental 
circumstances where {\it distances} rather than {\it times} are measured.
This issue can be studied 
in {\it single} neutral kaon production in reactions 
like $ \pi^- p \to \Lambda \k0$,  
$K^+ n \to p \k0$, $p\bar{p} \to K^+\pi^-\kb0$,
or in kaon pair production in reactions such as
$e^+e^- \to K_S K_L $ at $\Phi$ factories 
like Daphne \cite{4,5,6}. We will concentrate in the present paper
on single kaon production --- a problem which finds its analogy  
in recent discussions on neutrino oscillations \cite{7,8} ---
and point out the differences to a $\Phi$-factory at
an appropriate place.

The methods and arguments given in \cite{1,2,3} for the {\it single} 
kaon oscillations
(for the analogous neutrino case, see \cite{7,8}) 
to justify a particular 
derivation of the spatial probability $P(r)$
are rather different from each other and 
so are partially the results.  In our opinion,
the origin of the discrepancies (in both, the derivation and the results) 
can  
be traced back to the absence of a clean quantum mechanical definition 
of a probability which depends only on the distance $r$. If the  
question of what is the probability to detect either a $\k0$ or a  
$\kb0$ at a distance $r$ from the kaon production point is a legitimate 
one, then a quantum mechanical definition should indeed exist.  
Sometimes the 
classical formula $r=v t$ is invoked to transform $P(t)$ into $P(r)$. However,  
strictly speaking, 
this procedure might not be appropriate in a quantum mechanical context  
due to the finite spread  
of the wave packets. Furthermore, in the neutral kaon system the existence 
of a different mass for each component $K_L$ and  $K_S$ implies different 
mean velocities.

From the point of view of quantum mechanics a probability has to be defined
in terms of the wave function $\Psi (x,t)$ which encodes all the
information of the system. This probability should then be constructed
according to our experimental requirements. 
A position dependent probability
$P(r)$ for propagating states can only be obtained
by integrating an appropriately defined quantity over the time variable.
Following the above arguments one concludes that the quantity to be
constructed in the first place in terms of $\Psi (x,t)$
should be a probability density in time (dependent on $x$ and $t$)
which corresponds to the current density.
Indeed a physical detector at a distance $r$  
from the production point measures the  
time integrated flux of probability flowing across its surface , i.e., 
\bqn 
P(r)=\int_A d \vec{A} \int dt \jxt 
\label{1} 
\eqn 
with 
\bqn 
\jxt=\frac{d P}{dt dA}\vec{n}, 
\label{2} 
\eqn 
where $A$ is the surface and $\vec{n}$ is a unit vector pointing outwards. 
 
The problem then is to find a definition for the current in the case of 
{\it mixed} 
states such as $\k0$ and $\kb0$. Note that in quantum mechanics 
the current $\vec{j}(\vec{x},t)$ is usually defined only for states 
of definite mass, nevertheless we will show that the corresponding
generalization
is possible in our case. 
In doing so we will restrict ourselves for simplicity to $1+1$ dimensions. 
Moreover, since the controversial point in connection with   
$P_{\k0 \to \k0} (r)$ and $P_{\k0 \to \kb0} (r)$ refers to neutral kaon
propagation in space--time and it 
has neither to do with 
relativistic effects nor with CP--violation in the neutral kaon system,
we can work in  {\it i}) a non relativistic limit for which 
the current take simple expressions, {\it ii}) a CP conserving theory, 
and {\it iii}) a situation with stable kaons, $\Gamma_S=\Gamma_L=0$. 
The last two points are consistent with the
Bell--Steinberger relation in the limiting case of
$\langle K_S | K_L \rangle =0$ \cite{bs}.
Note that  
the non-relativistic limit refers to retaining the rest mass in the 
energy--momentum relation,
while neglecting for the moment higher orders of $v/c$. 
The modified Schr\"odinger equation is then: 
\bqn 
i \frac{\partial}{\partial t} \Psi_{S/L}(x,t)= \frac{-1}{2 m_{S/L}} 
\frac{\partial^2}{\partial x^2}\Psi_{S/L}(x,t)  + m_{S/L}\Psi_{S/L}(x,t), 
\label{3} 
\eqn 
which is the appropriate non-relativistic limit of either Klein-Gordon or 
Dirac equations. The relativistic counterparts including finite width effects
($\Gamma_{S/L} \neq 0$) 
will be easily inferred from our results. 
 
In next section we recall some basic formulae of the neutral kaon 
states and  of wave packets. In section 3 the current density is  
obtained and in section 4 we compute the oscillation probabilities. 
Some brief conclusions follow. 
 
\section{Wave packets for $\k0$ and $\kb0$} 
 
In the neutral kaon system one can distinguish two sets   
of states. The ``strong--interaction" basis is given by   
the states $\mid \k0 \rangle$ and $\mid \kb0 \rangle$,  
with well defined strangeness but undefined masses,  
fulfilling the orthogonality condition $\langle \k0\mid   
\kb0 \rangle = 0$. In contrast, the ``free--space" propagating states   
$\mid \ks \rangle$ and $\mid \kl \rangle$ 
have well defined masses,
$m_{S/L}$, 
but are not strangeness eigenstates.   
This is the appropriate set to describe time evolution in   
free space given by $|K_{S/L}(t)\rangle =e^{-i\lambda_{S/L}t}|K_{S/L}\rangle$
with $\lambda_{S/L} \equiv m_{S/L}-i/2\Gamma_{S/L} \to m_{S/L}$ in our limit. 
The relation between both sets is given by  
\bqn  
\mid K_{S/L}\rangle = p \mid \k0 \rangle \pm q \mid \kb0 \rangle  
\label{4}  
\eqn  
with $\mid p \mid ^2 +\mid q \mid ^2 =1$ and   
$\mid p \mid ^2 -\mid q \mid ^2 \neq 0$ if CP is violated in the mixing.  
  
A general space--time dependent state  $|K_{S/L}(x,t)\rangle$ 
can be written as  
\bqn  
\mid K_{S/L}(x,t)\rangle =  \Psi_{S/L}(x,t) \mid K_{S/L}\rangle  .
\label{5}  
\eqn  
From Eqs. (\ref{4})-(\ref{5}) one can now obtain the time evolution of
an initially produced $\k0$.
In the two dimensional space spanned by 
$\mid \k0 \rangle =\left(\begin{array}{c}  
1 \\ 0 \end{array} \right)$ and $\mid \kb0 \rangle =\left(\begin{array}{c}  
0 \\ 1 \end{array} \right)$ one has  
\bqna  
\mid\k0 (x,t)\rangle=\left( \begin{array}{c} \Psi_{\k0}(x,t) \\  
			\Psi_{\kb0}(x,t)  
	    \end{array} \right)& =&  
 \frac{1}{2}\left( \begin{array}{c} \Psi_{S}(x,t)+\Psi_{L}(x,t) \\  
			\frac{q}{p} \left( \Psi_{S}(x,t)-\Psi_{L}(x,t) \right)  
		   \end{array} \right)  \nn \\ 
&\longrightarrow& \frac{1}{2}\left( \begin{array}{c}   
  \Psi_{S}(x,t)+\Psi_{L}(x,t) \\      
  \Psi_{S}(x,t)-\Psi_{L}(x,t)  
		   \end{array} \right) ,  
\label{6}  
\eqna     
where the last term in~(\ref{6}) holds only in the  
CP conserving limit $p=q=1/\sqrt{2}$. In case of an  
initially produced $\bar{K^0}$ one similarly has
$\mid \kb0 (x,t)\rangle \to 1/2\left( \begin{array}{c} 
            \Psi_S -\Psi_L \\ -\Psi_S - \Psi_L 
            \end{array} \right)$.

The wave functions $\Psi_{S/L}(x,t)$ are taken to be wave packets which
are known to give a realistic description of the space--time evolution.   
We choose  
for simplicity a Gaussian form 
\bqn 
\Psi_{S/L}(x,t)=\frac{\sqrt{a}}{(2\pi)^{3/4}}\int_{-\infty}^{\infty}  
d k e^{-\frac{a^2}{4} (k-k_{S/L})^2} 
e^{i(k x -\omega_{S/L}(k) t)} ,
\label{7}  
\eqn  
where according to (\ref{3}) 
\bqn  
 E_{S/L}=\omega_{S/L}(k)=m_{S/L} + \frac{k^2}{2 m_{S/L}} , 
\label{8}  
\eqn   
and $k_{S/L}$ are the mean values of the momenta of the short/long
kaon states. Recall also that $a$   
is related to the width of  
the packet, $\Delta x =a/2$, and that the uncertainty relation   
gives $\Delta k \Delta x = 1/2$, i.e., $\Delta k=1/a$ for both $K_S$ and
$K_L$.  
Upon integration over $k$ an explicit formula for $\Psi_{S/L}(x,t)$ can be
obtained
\bqn 
\Psi_{S/L}(x,t)= \left(\frac{2}{\pi a^2}\right)^{1/4}  
\frac{1}{(1+\gamma_{S/L}^2 t^2 )^{1/4}} 
exp\left\{- \frac{(x-v_{S/L}t)^2}{a^2(1+\gamma_{S/L}^2 t^2)}\right\}
e^{i\varphi_{S/L}}, 
\label{9} 
\eqn 
with  
\bqn
\varphi_i=-\theta_i + k_i x - \omega_i (k_i) t +  
\frac{(x-v_i t)^2 \gamma_i t}{a^2 (1+\gamma_i^2 t^2)}, \ \ \ \\ \ \ i=S,L
\label{10} 
\eqn
$\omega(k)$ defined in Eq. (\ref{8}), and 
\bqn 
\gamma_i \equiv \frac{2}{m_i a^2} ,  \ \ \ \ \  
 v_i\equiv \frac{k_i}{m_i} ,  \ \ \ \ \ 
 \tan(2\theta_i)\equiv \gamma_i t . 
\label{11} 
\eqn 
Eqs. (\ref{9})-(\ref{11}) represent the usual Gaussian wave packet displaying a 
well--known spread in time for definite mass eigenstates $K_S$ or $K_L$.

Note that in our single $K^0$ production processes, energy--momentum 
conservation
requires that the two $K_{S/L}$ components of the $K^0$ wave packet should have
slightly different mean energies $E_{S/L}$ {\it and} momenta $k_{S/L}$ due to
the mass difference
$\delta m=m_L -m_S > 0$. In this sense we fully agree with the main point
raised by Srivastava et al.\cite{1}, also advocated in \cite{7}, concerning the
presence of two different energies and momenta. 
Both mass eigenstate components have to be produced in
the coherent superposition $1/2[\Psi_S (x,t=0)+\Psi_L (x,t=0)]$ corresponding to
an initial $K^0$. This coherent production of overlapping wave packets
of $K_S$ and $K_L$ also requires
\bqn
\label{ex1}
\Delta k >> \delta k=k_L-k_S 
\eqn
In this respect the situation is very similar to that considered 
in neutrino oscillations 
whose wave packet treatment has been discussed in \cite{8}. 
It is also
interesting to observe that in our treatment 
single $K^0$ production somehow mimics a
two--slit experiment in momentum space, $\delta m$ playing
the role of the ``separation"
between the $K_S$ and $K_L$ ``slits".

There is a subtle point worth mentioning here referring to the 
Bargmann superselection rule \cite{9},\cite{10}. This 
rule states that in non--relativistic quantum mechanics   
Galilean invariance forbids a coherent superposition of states with 
different masses. We do not consider this a serious problem here.
Beyond the non-relativistic limit 
such restrictions do not hold as is clearly experienced in the neutral 
kaon system. 
Note that Eq. (\ref{3}) is  a limit of a relativistic 
equation and that we retain the different rest masses. A relativistic
current has then to reduce to its non--relativistic counterpart 
which is based
on Eq. (\ref{3}). Notice also that 
Galilean invariance cannot be strictly maintained if, for instance, we 
would use time dependent potentials  in the Schr\"odinger 
equation (like space-time dependent electromagnetic potentials) \cite{10}.

\section{Currents for $ \k0 $ and $ \kb0 $ } 
 
The current density for the $K_{S/L}$ mass eigenstates whose wave functions
obey Eq. (\ref{3})
is defined as 
\bqn 
j_{S/L}(x,t)=\frac{1}{m_{S/L}} Im 
 \left( \Psi_{S/L}^*(x,t) \frac{\partial}{\partial x}\Psi_{S/L}(x,t) \right),
\label{12}  
\eqn 
and satisfies the usual continuity equation  
$\frac{\partial}{\partial t} \rho_{S/L} + \frac{\partial}{\partial x} j_{S/L} 
=0$, with $\rho_{S/L}(x,t)= \Psi_{S/L}^* \Psi_{S/L}$.  
For the $\k0$ and $\kb0$ states we expect that the corresponding 
currents are not conserved  
as these states mix in time  
due to the mass difference $\delta m$ in their $K_S$ and $K_L$ components. 
Thus, the modified continuity equation should read
\bqn 
\frac{\partial}{\partial t} \rho_{\k0,\kb0} + \frac{\partial}{\partial x}  
j_{\k0,\kb0} 
=d_{\k0,\kb0}, 
\label{13} 
\eqn 
where $\rho_{\k0}=\Psi^*_{\k0}\Psi_{\k0}$, 
$\rho_{\kb0}=\Psi^*_{\kb0}\Psi_{\kb0}$
and $d_{\k0,\kb0}$
is a kind of  
``diffusion term" to be determined. 
      
Let us list some general requirements for the current density  
$j_{\k0,\kb0}$: {\it i}) It should contain only ``velocity terms", i.e., 
terms of 
the form $ 1/m_i \left( \Psi_j^* \frac{\partial}{\partial x}\Psi_k \right)$, 
with $i,j,k=S,L$, whereas one has no similar condition for $d_{\k0,\kb0}$. 
{\it ii}) As the mass difference goes to zero, 
$\delta m=m_L-m_S \to 0$, we should have 
 $d_{\k0,\kb0}=0$, independently of the form of the wave packets $\Psi_{S/L}$. 
In this limit obviously one has $j_{\k0}=j_{\kb0}=j_{S}=j_{L}$. {\it iii}) 
In the zero decay width limit the  
total number of neutral kaons 
must be a conserved quantity, i.e., 
\bqn 
\frac{\partial}{\partial t} \left( \rho_{\k0}+\rho_{\kb0}\right) 
 + \frac{\partial}{\partial x}\left(  
j_{\k0} +j_{\kb0} \right)=0 
\label{14} 
\eqn 
Hence, $d_{K^0}=-d_{\bar{K^0}}$ should hold.
 
We can now derive the expression for the current density $j_{\k0,\kb0}$. 
First, we take the time derivative of the densities  
$\frac{\partial}{\partial t}\left( \Psi^*_{\k0,\kb0}\Psi_{\k0,\kb0}\right)$, 
 where 
the wave functions are given in Eq.~(\ref{6}). We then use the Schr\"odinger  
equation (\ref{3}) and separate the "velocity 
terms" according to point {\it i}) above. The linear combination of such 
terms will be called $j_{\k0,\kb0}$. All other terms that are not of  
velocity type are included in $d_{\k0,\kb0}$. 
We thus obtain 
\bqn 
j_{\k0,\kb0}= \frac{1}{4}\left(j_S+j_L \right)+ \frac{1}{2} j^{int}_{\k0,\kb0}, 
\label{15} 
\eqn 
where $j_{S/L}$ are the usual 
current densities (\ref{12}) corresponding to the $K_{S/L}$ states, 
and $j^{int}_{\k0,\kb0}$ contains the interference terms 
\bqn
j_{\k0}^{int}= \frac{1}{2} \left[  
\frac{1}{m_S} Im \left(\Psi_S^*\frac{\partial}{\partial x}\Psi_L \right)+ 
\frac{1}{m_L} Im \left(\Psi_L^*\frac{\partial}{\partial x}\Psi_S \right)   
 \right]=-j_{\kb0}^{int}.
\label{16} 
\eqn
For the ``diffusion terms" we get 
\bqn
d_{\k0}= \frac{1}{4}\left(\frac{1}{m_S}-\frac{1}{m_L}\right) 
	    Im (\frac{\partial}{\partial x}\Psi^*_L  
		     \frac{\partial}{\partial x}\Psi_S ) + 
	    \frac{1}{2}(m_L - m_S) Im (\Psi^*_S \Psi_L )= -d_{\kb0} 
\label{17} 
\eqn 
 
It is now easy to check that the three 
aforementioned requirements on the current densities 
are indeed fulfilled. In particular, $d_{\k0,\kb0} \to 0$ as  
$\delta m \to 0$, and Eq.~(\ref{14}) 
is also trivially satisfied due to the 
antisymmetry properties of $j_{K^0,\bar{K^0}}^{int}$ and $d_{\k0,\kb0}$ in
(\ref{16}) and (\ref{17}), respectively.

The current densities (\ref{15}) can also be written in terms of the wave 
functions $\Psi_{\k0}$ and $\Psi_{\kb0}$ 
\bqna 
j_{\k0}&=&\frac{1}{2}\frac{1}{m^2-{1 \over 4}\delta m^2}\left(  
	 2m Im (\Psi^*_{\k0}\frac{\partial}{\partial x}\Psi_{\k0})- 
	 \delta m  \,Im (\Psi^*_{\k0}\frac{\partial}{\partial x}\Psi_{\kb0}) 
	  \right) \nn \\ 
 j_{\kb0}&=&\frac{1}{2}\frac{1}{m^2-{1 \over 4}\delta m^2}\left(  
	 2m  Im (\Psi^*_{\kb0}\frac{\partial}{\partial x}\Psi_{\kb0})- 
	 \delta m \, Im (\Psi^*_{\kb0}\frac{\partial}{\partial x}\Psi_{\k0}) 
	  \right),
\label{18}	   
\eqna
where $m$ is the mean value of the mass, $m=(m_S+m_L)/2$.	  
Note that in contrast to the densities $\rho_{\k0,\kb0}$ each of the  
two currents $j_{\k0,\kb0}$ contains both wave functions  
$\Psi_{\k0,\kb0}$. This means that the constructed currents are unique in the
sense that all ``velocity terms" are contained in $j_{K^0, \bar{K^0}}$ and not in
$d_{K^0, \bar{K^0}}$.

\section{Distance Dependent Probabilities $P_{\k0 \to \k0 ,\kb0} (r)$ } 
 
Given the currents $j_{\k0,\kb0}$ in (\ref{15}) 
the $x$--dependent probabilities 
\bqna 
P_{\k0 \to \k0}  (x)&=&  \int_{0}^{\infty} dt \, j_{\k0}(x,t), \nn \\ 
P_{\k0 \to \kb0} (x)&=&  \int_{0}^{\infty} dt \, j_{\kb0}(x,t) 
\label{19} 
\eqna 
can now be calculated in analogy to Eq.~(\ref{1}) 
(recall that for simplicity we are 
working in 1+1 dimensions). Before evaluating the integrals of 
Eq.~(\ref{19}) it is convenient to elaborate on the kinematics  
involved. Note that we will use consistently the  
non--relativistic limit, Eq. (\ref{8}), and restrict ourselves
to the case of single kaon production in ``two--to--two" reactions like
$\pi^- p \to \Lambda K^0$, $K^+ n \to K^0 p$, etc. Working in the CM--system
with the total energy $\sqrt{s}=m+M+Q$, where $M$ is the final baryon mass, one
has  
\bqna 
\delta k\equiv k_L-k_S , \ \ \ \ \ \ \ \ \ \  
k \equiv \frac{1}{2}(k_S+k_L) \nn \\ 
\delta E\equiv E_L-E_S \simeq \frac{k\delta k}{m}+ 
\delta m \left(1-\frac{k^2}{2 m^2}\right), \ \ \ \ \  
E\equiv \frac{1}{2}(E_L+E_S) \nn \\
\delta v \equiv v_L - v_S \simeq {\delta k \over m}
-{k\delta m \over m^2}, \ \ \ \ \ 
v\equiv {1 \over 2}(v_S + v_L)  
\label{20} 
\eqna
The $Q$--value of the reaction can be chosen in the very wide range
\bqn
\label{21}
m >> Q >> \delta m
\eqn
imposed by our non--relativistic treatment and the need to produce a 
$K^0$ in a coherent $K_{S/L}$ superposition. Expanding the relevant variables
one finds
\bqna
k^2 &\simeq & {2Mm \over M+m}Q, \ \ \ 
k\delta k\simeq -\delta m
\left[{Mm \over M+m}-\left({M \over M+m}\right)^2Q\right]
\nn \\
E&\simeq &m + {m \over M+m}Q, \ \ \ \delta E \simeq \delta m
\left[1 - {M \over M+m}-{ M \over (M+m)^2}Q \right]
\label{22}
\eqna
and, more importantly,
%
\bqn
\frac{\delta v}{v} \simeq -{\delta m \over m}\left[{M+2m \over M+m}
+ {m \over Q}\right]\sim  
O\left( \frac{\delta m}{Q}\right)
\sim
O\left( \frac{\delta m}{m v^2}\right).
\label{23}
\eqn 
We stress that in spite of working in the non--relativistic limit
$v$ cannot be arbitrarily small as to imply a $Q$--value (or a
kaon kinetic energy)
of the order of the mass difference $\delta m$. 
In such unrealistic situation one 
would be favouring the production of the $K_S$ component over the 
slighlty heavier $K_L$ one.
Taking into account that $\delta m/m \sim 10^{-15}$ Eqs. (\ref{21}) and
(\ref{23}) imply $v>>10^{-7}$.
Notice, however, that even velocities of the order
of $10^{-3}$ would require an unrealistic high degree of fine tuning of the
CMS energy $\sqrt{s}$ ($Q \sim keV$).


At this stage it is instructive to point out some differences to
a $\Phi$-factory like Daphne. In the latter case one usually studies  
correlation probabilities of $K_S$ and $K_L$ {\it decays}
depending on two space-time
points \cite{4}. 
The kinematics dictates  
in this case $\delta v/v=-(\delta m/m)$ rather than (\ref{23}). 
 
In terms of the Gaussian wave packets (\ref{9}), 
the currents $j_{S/L}$ appearing in (\ref{15}) can be cast into a simple form 
\bqn 
j_i= \sqrt{\frac{2}{\pi}}\frac{v_i + x \gamma_i^2 t}{a(1+\gamma_i^2 t^2)^{3/2}} 
\exp{\left[ -2 \frac{(x-v_i t)^2}{a^2(1+\gamma_i^2 t^2)}\right]}, 
 \ \ \ \ \ \ \  i=L,S
\label{24} 
\eqn 
Introducing the new variables 
\bqn 
a z_i \equiv \frac{x-v_i t}{(1+\gamma_i^2 t^2)^{1/2}} 
\label{25} 
\eqn 
we can compute to a very good approximation 
\bqn 
\int_{0}^{\infty} dt \, j_i (x,t) = \sqrt{\frac{2}{\pi}}  
\int_{-\frac{1}{2} a k_i<<-1}^{\frac{x}{a}>>1} d z_i\, e^{-2 z_i^2} \simeq 1 
\label{26} 
\eqn 
 
Indeed in the upper limit $x>>a$ since the position $x$ of the 
detector is considered 
to be far from the $K^0$ production point, 
far in comparison to the width of the
packet $a/2$. Otherwise, one would be measuring the inner wave packet effects
rather than its space--time propagation 
which is our interest here. 
In the lower limit of the integral
the momentum $k$ is much larger than its dispersion
$\Delta k=1/a$ for any realistic velocity $v\sim 10^{-3}-10^{-1}$.

Neglecting in (\ref{16})  polynomial terms of order $\delta m/m$ and expanding
further the remaining expression in leading order of $\delta m/m$, we can write
\bqna
j^{int}_{K^0} &\simeq& \sqrt{{2 \over \pi}}
{v + x\gamma^2 t \over a(1 + \gamma^2
t^2)^{3/2}}exp \left\{-2{(x-vt)^2 \over a^2(1+ \gamma^2t^2)}\right\}
exp\left\{-{t^2\delta v^2 \over 2a^2(1+\gamma^2 t^2)}\right\} \nn \\
&\times& \left[\cos \theta (x,t) + \sin \theta (x,t){\delta v \gamma t
\over 2(v+x\gamma^2 t)}\right]
\label{27}
\eqna
where
\bqn
\theta (x,t)
\equiv -\delta Et + \delta kx + {2 t\delta v (vt-x) \over a^2}
{\gamma t \over (1+\gamma^2 t^2)}
\label{28}
\eqn
is the crucial phase 
and the mean $\gamma \equiv 2/ma^2$, $v \equiv k/m$ have been
defined in analogy to Eq.~(\ref{11}). 
Observing the similarities between the
Eqs. (\ref{24}) and (\ref{27}),
we introduce a variable as in (\ref{25}), $az={x-vt \over (1 +\gamma^2
t^2)^{1/2}}$,  
whose inverse in the region of interest $0 \leq t \leq \infty$ is 
\bqn 
t(z)=\frac{vx - a z\sqrt{v^2 + \gamma^2(x^2-a^2 z^2)}}{(v^2-\gamma^2 a^2 z^2)} 
\label{29} 
\eqn 
for $v>0$ and $x\geq 0$. The integrand has now a form of 
$e^{-2 z^2} f(z)$ which 
can be expanded in a Taylor series around $z=0$. Note that $t(0)=x/v$ and that
the coefficient of $\sin \theta $ in (\ref{27}) is bounded by
$1/2 (\delta v/v){\gamma t(0) \over 1 + \gamma^2 t^2(0)} \leq 1/4(\delta 
v /v) \sim O(\delta m/Q)$ and hence small by virtue of Eq. (\ref{21}).
In leading order we then obtain
\bqna 
\int_{0}^{\infty} dt \, j^{int}_{\k0}(x,t) &\simeq& \sqrt{{2 \over  \pi}}
\cos \theta (x,t(0))
exp\left\{-{t^2(0) \delta v^2 \over 2a^2(1 +\gamma^2 t^2(0))v^2}\right\} 
\int_{-\infty}^{\infty}dz e^{-2z^2}\nn \\
&\simeq& \cos \theta (x, x/v) 
\label{30} 
\eqna 
since the additional Gaussian factor has a negligible exponent,
$1/2(\delta v^2/\gamma^2){\gamma^2 x^2/v^2 \over 1+ \gamma^2x^2/v^2}
\leq 1/2(\delta v^2/\gamma^2) \sim 1/2(\delta k/\Delta k) << 1$ 
due to (\ref{ex1}).

Taking into account the kinematics given in Eqs. (\ref{21})-(\ref{23})
 the phase turn out to be
\bqn
\label{31}
\theta (x, x/v)=-\delta m \left (1 -{1 \over 2}v^2 \right ){x \over v}.
\eqn
Putting now everything together we obtain
 \bqn 
P_{\k0 \to \k0,\kb0}(x) =  
	       \frac{1}{2} \left( 1 \pm  
	      \cos\left[ \delta m \frac{x}{v}(1-\frac{1}{2} v^2) \right] 
	      \right) 
\label{32} 
\eqn 
which is our final result in non--relativistic approximation and for stable
kaons.
Reintroducing the non--vanishing $\Gamma_{S/L}$ decay widths and higher orders
in $v/c$,
Eq. (\ref{32}) can be cast into its final form as
\bqn
P_{\k0 \to \k0, \kb0}(\tau) =  
	      \frac{1}{4} \left( e^{-\Gamma_S\tau} + e^{-\Gamma_L\tau}\pm
	      e^{-{1 \over 2}(\Gamma_S+\Gamma_L)}\cos[\delta m \tau]\right),
\label{33} 
\eqn 
which corresponds to the usual expression in terms of the proper time $\tau$. 
%

\section{Conclusions} 
We have argued that in order to avoid theoretical ambiguities 
it is essential to use a proper definition of space--dependent
probability $P(r)$: the 
time and surface integrated current density, $j(\vec{x},t)$. 
We have shown that it 
is possible to construct such a current density for coherent superposition
of different mass eigenstates. 
Further, we have used throughout wave packets 
which describe the physical space--time evolution and take into account the  
necessary overlapping of wave functions to obtain interference  
effects. Wave packets have also been used in \cite{4} to compute
space--dependent decay correlation probabilities at a $\Phi$--factory. Their
results are quite in line with those obtained in the present paper.

In our treatment the two components $\ks$ and $\kl$ of the neutral kaon  
evolve in space--time with different momenta and energies 
as dictated by energy--momentum conservation 
in a situation that can be visualized as a two slit experiment in 
momentum space.
We have then obtained the oscillatory term $\cos\left[ 
\delta m \tau \right]$ , which corresponds to the usual expression but
without the recourse to any classical formula to transform time probabilities  
into space probabilities. 

\subsection*{Acknowledgments} 
This paper has been partially supported by CICYT contracts AEN94--936 and
AEN95--815
and EURODAPHNE, HCHP, EEC
contract no. CHRX-CT 920026.
R.M.T.\ acknowledges a EC post-doctoral fellowship ERBC HBCICT--941777. 
M.N. thanks the Spanish Ministerio de Educacion y Ciencies for a post--doctoral
fellowship.

\end{document}